\documentclass[final,3p,times,twocolumn]{elsarticle}

\usepackage{graphicx}
\usepackage{amssymb}

\newcommand{\bra}[1]{\langle #1|}
\newcommand{\ket}[1]{|#1 \rangle}
\newcommand{\vl}{|}
\newcommand {\e}{{\rm e}}

\newcommand{\Hs}{\mathcal{H}_{\rm s}}

\newcommand{\kB}{k_{{\rm B}}}

\newcommand{\g}{g}
\newcommand{\EJ}{{E_{\rm J}}}

\newcommand{\rhoqc}{\rho_{\rm q-c}}

\newcommand{\up}{{\uparrow}}
\newcommand{\down}{{\downarrow}}
\newcommand{\Pud}{P_{{\uparrow} \to {\downarrow}}}

\newcommand{\vel}{v}

\journal{Physica E}
\begin{document}

\begin{frontmatter}

\title{Entanglement and Disentanglement in Circuit QED Architectures}

\author{David Zueco, Georg M. Reuther, Peter H\"anggi, and Sigmund Kohler} 
\address{Institut f\"ur Physik, Universit\"at Augsburg,
         Universit\"atsstra{\ss}e~1, D-86135 Augsburg, Germany}

\begin{abstract}
We propose a protocol for creating entanglement within a dissipative
circuit QED network architecture that consists of two electromagnetic 
circuits (cavities) and two superconducting qubits. The system
interacts with a quantum environment, giving rise to decoherence and
dissipation.  We discuss the preparation of two separate entangled
cavity-qubit states via Landau-Zener sweeps, after which the cavities
interact via a tunable ``quantum  switch'' which is realized with an
ancilla qubit. Moreover, we discuss the decay of the resulting
entangled two-cavity state due to the influence of the environment,
where we focus on the entanglement decay.

\end{abstract}


\end{frontmatter}


\section{Introduction}
\label{intro}

The coupling between an atom and the quantized electromagnetic field 
represents a paradigmatic setup in quantum optics that allows one to 
study the interaction of light and matter in a fully quantum way. 
In most related experiments, an atom is placed inside a cavity, from
where stems the labelling ``Cavity Quantum Electrodynamics''
\cite{Walther2006a}. 

In the last years the implementation of such systems in the solid
state realm has been demonstrated \cite{Wallraff2004a,Hennessy2007a}.
Depending on the experimental implementation, the atom is realized
artificially with a superconducting circuit or a quantum dot, while a  
quantum LC-circuit as part of a transmission line or a photonic
crystal, respectively, serves as  cavity. This not only provides  a widely 
enhanced  tunability  of  parameters in an experiment, but also
allows one to  achieve the strong  coupling  limit  between the
artificial  atom  and the cavity, where the interaction exceeds the
decay rates of the atom and the cavity by far. It has recently been
shown  that  the coupling strength may even be enhanced by several orders
of  magnitude compared  to conventional cavity QED experiments
\cite{Blais2004a}. 
Furthermore, circuit QED systems may be used for many experimental
validations of quantum mechanics, such as the
quantum-non-demolition-like readout of  a qubit state
\cite{Lupascu2007a}, the generation of Fock states
\cite{Hofheinz2007a}, the observation of Berry phases
\cite{Leek2007a}, or multiphoton resonances
\cite{Deppe2008a}. Generation of entanglement between the qubit and the
cavity via a Landau Zener sweep has been theoretically proposed as well
\cite{Wubs2007a,Saito2006a}.  
From a fundamental point of view these setups, realizing ``quantum
optics on a chip'', give the opportunity to rediscover and improve the
vast goals of quantum optics and to test  experimentally its
concepts  in the solid state.

However, the mentioned experiments involve ``only'' a single qubit
and one cavity. With a view to quantum computing,
i.e.\ in order to realize a quantum register, it is highly
desirable to couple several of these qubit-cavity systems to
obtain a circuit QED network. This would not only allow for performing
quantum operations such as quantum gates, but also enable the
creation and distribution of entanglement throughout the network.
Indeed, a setup to couple two cavities via an ancilla qubit  has been
recently proposed in \cite{Mariantoni2008a, Helmer2007a}, 
while in recent experiments \cite{Sillanpaa2007a, Majer2007a},
two qubits have been entangled via one cavity.

In this paper, referring to the setup of Ref.~\cite{Mariantoni2008a}
we suggest the  transfer of entanglement between two qubit-cavity
systems via an ancilla qubit, which requires entangling different 
cavities. We restrict ourselves to the case where the  
qubit-cavity entanglement is created via a Landau-Zener sweep;
cf.\ Ref.~\cite{Wubs2007a,Saito2006a}.

Due to the interaction  with a dissipative environment, entanglement
within an open quantum system is typically subject to decay via
spontaneous emission. For the non-trivial class of states discussed
here, this can even occur during finite time. This phenomena of
``sudden death of entanglement'' has been investigated theoretically
for a conventional cavity QED setup \cite{Yu2002a} and observed in an
experiment as well \cite{Almeida2007}.

The paper is organized as follows. In section
\ref{entanglement-lzsweep} we shortly review the elementary
description of a superconducting qubit coupled to an electromagnetic
quantum circuit. Further, we give a brief review on the creation of
entanglement between qubit and cavity by a Landau-Zener sweep. In
section \ref{dissipative} we discuss the influence of a dissipative
environment on the final entangled qubit-cavity state. In section
\ref{sec:quantumnetworks} we consider a network architecture
consisting of two qubit-cavity systems dynamically coupled via an 
ancilla qubit, and discuss how this  dynamical interaction yields an 
entangled two-cavity state. Finally, in section \ref{suddendeath}, we
investigate the finite-time disentanglement of that state under
environmental influence, depending on the details of state preparation.


\section{Qubit-cavity entanglement via LZ sweep}
\label{entanglement-lzsweep}

Among many possible setups for circuit QED, one concrete realization
is given by a  Cooper pair box (CPB) that couples capacitively to a
LC circuit, which acts as a harmonic oscillator
\cite{Blais2004a}.  The CPB is formed of a superconducting island
connected to a superconducting reservoir by a dc SQUID. 
The effective Josephson energy $E_\mathrm{J} = E_\mathrm{J}^0
\cos(\pi\Phi/\Phi_0)$ can be tuned via an external flux $\Phi$
penetrating the SQUID, where
$\Phi_0$ denotes the flux quantum.  We assume that $\Phi$ is 
switchable within a sufficiently large time interval such that 
$E_\mathrm{J}{=} \hbar vt$, $v\,{>}0$.  The capacitive energy of the CPB, 
$E_\mathrm{el} {=} \frac{1}{2} E_\mathrm{c} (N-N_g)^2$ is 
determined by the material-dependent charging energy $E_\mathrm{c}$, 
the number $N$ of Cooper pairs on the island, and the
reduced background charge $N_g$ which is proportional to a voltage bias
$V_g$. In the charging limit $E_\mathrm{c} {\gg} \,E_\mathrm{J}$ and near
the charge degeneracy point $N_g{=}1/2$, it is justified to describe the
CPB by its two lowest charge states $|N{=}0\rangle$ and $|N{=}1\rangle$.
This allows the interpretation as a two-level system (TLS) or qubit
for which the Hamiltonian pseudo-spin notation reads $H_\mathrm{qb}
{=} -\frac{1}{2} \hbar vt \sigma_z$. It possesses the eigenstates
$|{\up}\rangle =  (|0\rangle+|1\rangle)/\sqrt{2}$ and $|{\down}\rangle
= ( |0\rangle -   |1\rangle)/\sqrt{2} $.  

Making use of the standard Rabi model the total qubit-oscillator
Hamiltonian reads \cite{Blais2004a}
\begin{equation}
\label{Rabi}
\Hs =
     -\frac{\hbar vt}{2}\sigma_z
     +\hbar \g\sigma_{x} ( a^{\dag} + a )
     +\hbar\Omega a^{\dag} a \, .
\end{equation}
The second term refers to the coupling of the qubit to the fundamental 
mode of the transmission line, which is modeled as a harmonic
oscillator with the usual bosonic creation and annihilation operators
$a^\dag$ and $a$ and energy eigenstates $|n\rangle$,
$n{=}0,1,\ldots,\infty$. Note that the usually performed Rotating Wave
Approximation (RWA) in the coupling Hamiltonian is not justified here
since during almost the complete Landau-Zener sweep, the qubit and the
harmonic oscillator are far detuned. 
%
\begin{figure}[tb]
\centerline{\includegraphics[scale=.75]{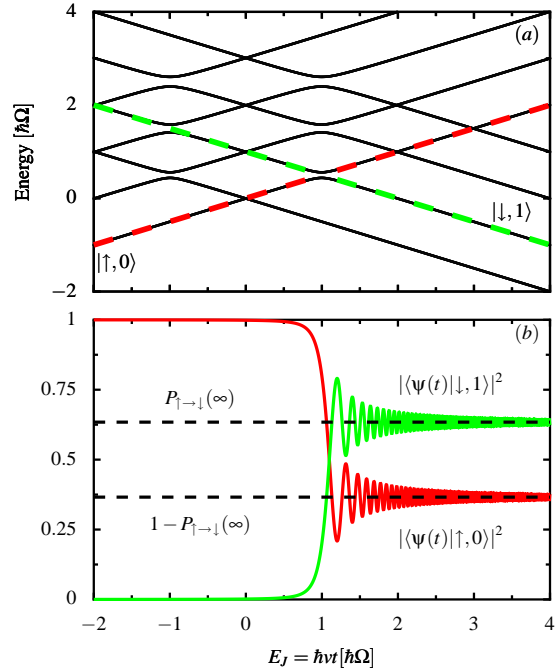}}
\caption{\label{fig:espectrum}  (Color online) (a) Adiabatic energy
  levels of the   qubit-oscillator Hamiltonian (\ref{Rabi}) as a
  function of   the Josephson energy which is swept at constant
  velocity such that   $\EJ= \hbar vt$. Here, $g=0.04 \Omega$ and
  $v=0.01\Omega^2$. (b) Dynamics for the probabilities 
  $\left| \langle {\down},1 | \psi (t) \rangle \right|^2$
  and  $\left|\langle {\up},0  | \psi (t) \rangle
  \right|^2$. The occupation of higher states is always less than
  $0.01\%$ (not shown). } 
\end{figure}
%

If the effective Josephson energy is switched from a large negative
to a large positive value, and assuming that both the TLS and the
oscillator are initially in their ground state
$|{\up},0\rangle$, the  probability for the TLS  to end up in the
upper state  can be computed exactly to read
\cite{Saito2006a}  
\begin{equation}
\label{PLZ}
\Pud = 1 - \e^{-2 \pi g^2/ v} .
\end{equation}
This clearly reminds one of the Landau-Zener formula for the bare
two-level system \cite{Landau1932a,Zener1932a,Stueckelberg1932a}, with
the level splitting in the anticrossing at $t{=}\Omega/v$ (see
Fig.~\ref{fig:espectrum}) now given by the coupling constant $g$. It
should be noted that the probability (\ref{PLZ}) does not provide
any direct information about the final oscillator state. However, due
to the symmetry of the Hamiltonian (\ref{Rabi}), every
creation or annihilation of a photon is accompanied by a qubit flip,
which restricts the resulting dynamics to the states $\ket{\up, 2n}$
and $\ket{\down, 2n+1}$. Furthermore, the ``no-go-up'' theorem
states that $P_{\up,n\to\up,m}=0$ for $m{>}n$ \cite{Saito2007a}. This
reduces the possible final qubit-oscillator states to the state
\begin{equation}
\label{finalestate}\hspace{-2em}
\ket{\psi (\infty)} = \sqrt{1-\Pud}\, \ket{{\up}, 0} 
+ \hspace{-.3em} \sqrt{\Pud} \sum_{n} c_{n} \ket{{\down}, 2n+1} 
\end{equation}
with the normalization $\sum_{n} \vl c_{n} \vl ^{2} =1$.
In the experimentally relevant limit of $\g \,{\ll} \,\Omega$, we
numerically find that during the whole Landau-Zener transition, the 
state is well approximated by 
\begin{equation}
\label{twostate}
\ket{\psi (t)} \cong \alpha (t) \ket{{\up}, 0} 
+ \beta(t) \ket{{\down}, 1} 
\end{equation}
with $\alpha(\infty)\, {=} \sqrt{1-\Pud}\,$, and $\beta (\infty) \,{=}
\sqrt{\Pud}\, c_{1}$, i.e.\ $\vl c_{1} \vl \cong 1 $. We substantiate this 
by plotting a typical example for  the characteristics of the LZ
dynamics in Fig.~\ref{fig:espectrum} (b), where it becomes evident
that finally only the states $ \ket{{\up}, 0}$ and $\ket{{\down}, 1} $ are
significantly populated. This means a great simplification since it allows
one to reduce the number of all relevant oscillator states to two. 

\begin{figure}[tb]
 \centerline{\includegraphics[scale=.75]{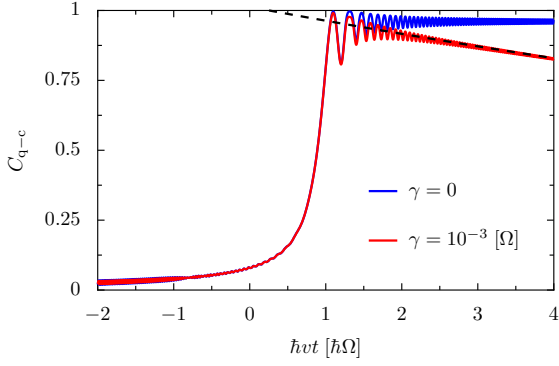}}
 \caption{\label{fig:concurrenceqc} (Color online) Qubit-oscillator
   entanglement in terms of the concurrence $C_\mathrm{q-c}$ for a
   Landau-Zener  transition. Here, $\kB T {=} 0.01 \hbar \Omega$, while
   $g$ and $v$ are  as in Fig.~\ref{fig:espectrum}. For finite
   coupling    strength $\gamma\,{>}\,0$ (red line), the decay of the
   concurrence is approximated by expression~(\ref{Cqc1}) (dashed
   black line). For comparison, the concurrence is also
   given for the case of vanishing interaction with the environment
   (blue line).}    
 \end{figure}

Then in particular, the qubit-oscillator state (\ref{twostate}) is
entangled  and can therefore be mapped to the state of two entangled
two-level systems. For future convenience it is useful to introduce
the density matrix for the qubit-cavity system $\rhoqc \,{=}
\ket{\psi} \bra{\psi}$ that reads, in the basis $\{\ket
{{\uparrow},0}, \; \ket{{\downarrow},0}, \; \ket {{\uparrow},1}, \;
\ket {{\downarrow},1} \}$, 
\begin{eqnarray}
\label{rhoclosed}
\rhoqc (t)=
\left (
\begin{array}{cccc}
\vl \alpha(t)\vl^{2} & 0 &0 & \alpha^* (t)\beta(t)
\\
0 & 0& 0& 0
\\
0 & 0& 0& 0
\\
\alpha (t)\beta ^*(t) &  0 &0 & \vert \beta(t) \vert^{2}
\end{array}
\right ) \, .
\end{eqnarray}
As a consequence, we can use the concurrence as a well-defined
measure of entanglement for the two two-level systems. It  is defined
as $C=\max \{\lambda_1-\lambda_2-\lambda_3-\lambda_4,0\}$. The parameters 
$\lambda$ are the ordered eigenvalues of the matrix $\sqrt {\rho}
(\sigma_y^{\rm {c}} \otimes \sigma_y^{\rm {q}}) \rho (\sigma_y^{\rm {c}} \otimes
\sigma_y^{\rm {q}})\sqrt {\rho}$ \cite{Wootters1998a}. Here, the Pauli matrices
$\sigma_y^{\rm {q}}$ and $\sigma_y^{\rm {c}}$ act on the qubit space
and the reduced oscillator space formed by the states $|0\rangle$ and
$|1\rangle$, respectively. For the density matrix
(\ref{rhoclosed}), the concurrence is given by
\begin{equation}
\label{cqc}
C_{{\rm q-c}} 
= 2 \vl \alpha^{*} (t)\beta (t) \vl \,.
\end{equation}
Thus, the amount of entanglement in the long-time limit 
is given by $C_{{\rm q-c}} (\infty) = 2\sqrt{(1-\Pud) \Pud}$, which
just depends on the ratio $\g^{2}/\vel$. As an example, we plot the
typical dynamics for the concurrence in Fig.~\ref{fig:concurrenceqc},
from where it is evident that entanglement is created after passing the
anticrossing.


\section{Influence of the dissipative environment}
\label{dissipative}

In a realistic experimental scenario, the qubit-oscillator system has
to be regarded as an open system, i.e.\ one that interacts with its
environment, thus suffering dissipation and decoherence.
Dissipative effects in an electromagnetic circuit are characterized by
a spectral density $I(\omega)$ which, within a quantum mechanical
description, can be modelled by coupling the circuit bi-linearly to
the field modes of an electromagnetic environment \cite{Yurke1984a}. 
For weak system-bath coupling, tracing out the bath degrees of freedom 
and following standard techniques, the bath can be eliminated within
Bloch-Redfield theory \cite{Redfield1957a, Blum1996a} yielding the
quantum master equation
\begin{eqnarray}
\label{QME-cqed}
\nonumber \hspace{-2em} \dot\rho & =&  -\frac{i}{\hbar}[\Hs,\rho ] 
 - \frac{1}{\hbar}   [ Q, [\hat Q, \rho] ] 
 - i\frac{Z_0}{\hbar} [ Q, [ \dot Q, \rho ]_+ ] \\
&\equiv&  -\frac{i}{\hbar}[\Hs,\rho ] + \mathcal L \rho ,
\end{eqnarray}
with the anticommutator  $[A,B]_+\, {=}\, AB\, {+}\, BA$, the
oscillator-bath coupling operator $Q = a^{\dag} {+}\, a$ and the
operators $\dot Q {=} i[\Hs,Q]/\hbar$ and 
\begin{equation}
\hat Q = \frac{1}{\pi} \int_0^\infty d\tau \int_0^\infty d\omega\,
   S(\omega) \cos(\omega\tau) \tilde Q(-\tau) .
\end{equation}
Here, $S(\omega) = I(\omega)\coth(\hbar\omega/2k_{\rm B} T)$ is the
Fourier transform of the symmetrically-ordered equilibrium
correlation function for a thermal environment at temperature $T$. We
assume the spectral density to be ohmic, i.e. $I(\omega) =
\omega \gamma$ with an effective impedance $\gamma$. 
 The notation $\tilde X(t)$ is a shorthand for the
Heisenberg operator $U_0^\dagger(t) X U_0(t)$, where $U_0$ is the
system propagator. While the quantum master equation (\ref{QME-cqed})
is general, an explicit form with respect to the Hamiltonian
(\ref{Rabi}), together with details about the numerical implementation
can be found in Ref.~\cite{Zueco2008a}.    

\subsection{Entanglement creation}
Due to the influence of the environment, in particular, the
entanglement created between qubit and cavity is subject to decay
during the Landau-Zener sweep. In the case of low temperatures, $\kB T
{\sim}\, 0.01 \hbar \Omega$, which is relevant for most experiments, our
numerical investigations of the master equation (\ref{QME-cqed}) confirm the
following scenario:
Before reaching the first avoided crossing at time
$t\,{<}\,{-}\,\Omega/v$ (see Fig.~\ref{fig:espectrum}), the system
remains in its ground state. Then at time $t\,{=}\,{-}\,\Omega/v$, it
evolves  into the superposition (\ref{twostate}). Simultaneously, the 
bath causes coherence decay and, for low temperature, a relaxation
from the state $\ket{{\up},1}$ to $ \ket{{\up},0}$ occurs.  We
emphasize here that there is no transition $\ket{{\up}, 0}\, {\to}
\,\ket{{\down},0}$ since the qubit experiences an effective heat bath
with a spectral density sharply peaked at the oscillator frequency
$\Omega$ \cite{Thorwart2000a, Thorwart2004a, Wilhelm2004a, Nesi2007a}.
Thus, for large times $t\gg\Omega/v$, the spectral density at the
qubit splitting $\hbar vt$ vanishes and, consequently, the qubit is
effectively decoupled from the bath. As a consequence, we numerically
find that after the anticrossing the state $\ket{{\down},0}$ becomes
populated in addition to the states $\ket{{\up}, 0}$ and
$\ket{{\down}, 1}$. Quantitatively, the corresponding density matrix
takes the form [cf. (\ref{rhoclosed})]
\begin{eqnarray}
\label{rhoopen}
\rhoqc (t)=
\left (
\begin{array}{cccc}
a & 0 &0 & c
\\
0 & z& 0& 0
\\
0 & 0& 0& 0
\\
c^{*} &  0 &0 & d
\end{array}
\right ) \, .
\end{eqnarray}
%
After the Landau-Zener transition, its entries can be well approximated by 
\begin{eqnarray}\label{coeffs}
\nonumber a &=& 1 - \Pud \\
\nonumber  d &\cong& 
\Pud {\rm e}^{-\gamma(t - \tau_\mathrm{LZ} )} \\
\nonumber  z &\cong&
\Pud \left(1 - {\rm e}^{-\gamma(t - \tau_\mathrm{LZ})}\right) \\
\vl c  \vl
&\cong&
 \sqrt{\Pud(1-\Pud)}\,{\rm e}^{-\gamma(t - \tau_\mathrm{LZ})}\, ,
\end{eqnarray}
where $\tau_\mathrm{LZ} =  \Omega/v$ and $\Pud$ has been defined in
Eq.~(\ref{PLZ}). For the state (\ref{rhoopen}), the concurrence is now
given by  
\begin{equation}
\label{Cqc1}
C_{{\rm q-c}} = 2 \vl c  \vl
\cong
2  \sqrt{\Pud(1-\Pud)}\, {\rm e}^{-\gamma(t - \tau_\mathrm{LZ})} .
\end{equation}
Hence the loss of entanglement is directly related to the loss
of coherence between the qubit and the cavity and exhibits a simple 
exponential decay, see Fig.~\ref{fig:concurrenceqc}. 



\section{Qubit-cavity networks}
\label{sec:quantumnetworks}
\begin{figure}[tb]
\centerline{\includegraphics[scale=.5]{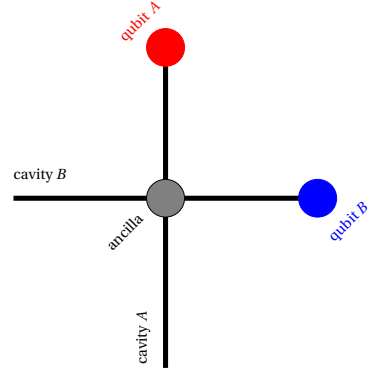}}
\caption{\label{fig:network}  (Color online) Sketch of the quantum
  network under   consideration. Qubits are represented by big dots
  and cavities by   black lines. Qubit $A$ ($B$) is only coupled to
  cavity $A$ ($B$),   whereas both cavities may effectively interact
  with each other by    the ancilla qubit.}   
\end{figure}

In this section we discuss how entanglement may be created within a
quantum network  architecture consisting of  several pairs of cavities
and qubits. We shall make use of the setup proposed in
Ref.~\cite{Helmer2007a} together with the state preparation via a
Landau-Zener sweep discussed above, in order to entangle two cavities.
We study the resulting dynamics in presence of dissipation and
decoherence.   
The setup we have in mind is sketched in Fig.~\ref{fig:network} and
consists of two qubit-cavity systems. The cross geometry indicates
that  qubit $A$ ($B$) is only coupled to cavity $A$ ($B$) whereas the
central qubit (ancilla) is coupled to  both cavities and
may be used as a switch, as we will explain below. 
The Hamiltonian of such a system can be approximated as
\begin{eqnarray}
\label{Hnetwork}
\hspace{-2em}\nonumber 
H & = & - \sum_{\mu=A,B} \Big \{ \frac{E_{J}^{\mu}(t)}{2}\sigma_{z}^{\mu}
 - \hbar g (a_{\mu}^\dagger+a_{\mu}^{\vphantom{\dagger}})\sigma^{\mu}_{x} + \Omega
 a_{\mu}^{\dagger} a_{\mu}^{\vphantom{\dagger}} \Big \} \\ 
\hspace{-2em}\nonumber && +
\frac{\omega_\mathrm{anc} (t)}{2}\,\sigma_{z}^{\rm anc} +
\left(\g_{x}^{\vphantom{\dagger}} 
\sigma_{x}^{\rm anc}+ g_{z}^{\vphantom{\dagger}} \sigma_{z}^{\rm anc}\right)
\sum_{i=1,2} (a_{j}^\dagger+a_{j}^{\vphantom{\dagger}})\\
\hspace{-2em}&&+\, G (a^{\dagger}_{A} + a_{A}^{\vphantom{\dagger}})
(a^{\dagger}_{B} + a_{B}^{\vphantom{\dagger}}) \; . 
\end{eqnarray}
The first line of Eq.~(\ref{Hnetwork}) denotes the usual Rabi
Hamiltonians for the respectively coupled qubits and cavities. For
simplicity, both cavities are supposed to have the same frequency
$\Omega$. The second line describes the ancilla qubit and its coupling
to the cavities, whereas the last line  accounts for the direct 
geometrical interaction between the cavities which will
be referred to as ``geometrical coupling'' \cite{Mariantoni2008a}.    

\subsection{The entanglement protocol}
In order to entangle the cavities $A$ and $B$, we pursue a ``quantum
switching protocol'' in analogy to the proceeding described in
Ref.~\cite{Mariantoni2008a}:
In order to reach the dispersive coupling regime $g_{x}/\Delta\, {\ll}
\,1$ between ancilla and cavities with the detuning $\Delta\, {=}\,
\omega_{\rm anc}\,{-}\, \Omega$, the ancilla qubit is widely detuned from
the cavity frequency \cite{Blais2004a}. On the other hand, $\Delta\,
{\ll}\, \Omega$, which allows a rotating-wave approximation with
respect to the ancilla-oscillator coupling.  Under this condition the
ancilla gives  rise to an effective
``dynamical coupling'' between the cavities which may be expressed as $
g_{x}^{2} \sigma_{z} (a_{A}^{\dagger} a_{B}^{\vphantom{\dagger}} 
+a_{A}^{\vphantom{\dagger}} a_{B}^{\dagger})/\Delta$. The total
interaction between both cavities including the additional geometrical
coupling [last line of (\ref{Hnetwork})]  is then given by   
\begin{equation}
\label{Hint}
\hspace{-1.5em}
H_{{\rm int}} = 
g_{{\rm SW}}
\;
(a_{A}^{\dagger} a_{B}^{\vphantom{\dagger}}
+a_{A}^{\vphantom{\dagger}} a_{B}^{\dagger}) \; ,
\quad
g_{{\rm SW}} =  \frac{g_{x}^{2}}{\Delta} \sigma_{z} + G .
\end{equation}
It is possible to ``switch off'' this interaction by setting  
$g_{{\rm SW}}$ to zero. This requirement can be fulfilled varying
$\Delta$ or $g_x$, or manipulating the state of the ancilla. 
\begin{figure}[tb]
\centerline{\includegraphics[scale=.75]{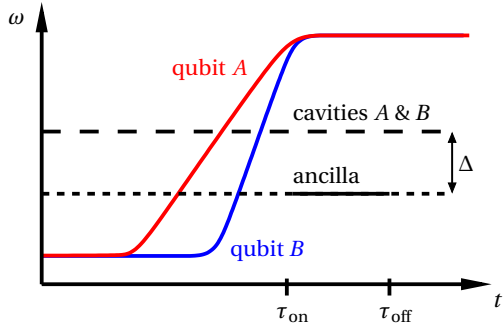}}
\caption{(Color online) Sketch of the quantum switch protocol,
  displaying the time-dependent qubit and cavity energies and the
  ancilla interaction: The qubits $A$ and $B$ undergo Landau-Zener
  sweeps with their respective cavities with velocities $v_A$ and
  $v_B$. This leads to the entangled qubit-cavity state
  (\ref{twostate}). Meanwhile, the cavities do not interact, i.~e. 
  $g_\mathrm{SW} = 0$ (dashed line). From $t=\tau_\mathrm{on}$ to
  $t=\tau_\mathrm{off}$,  $g_\mathrm{SW} > 0$  (solid line),
  leading to the entangled state (\ref{rhocav1}) which is subject to 
  decay for $t>\tau_\mathrm{off}$. For further details, see text.} 
\label{fig:protocol}  
\end{figure}
The protocol is now performed as follows: At first stage, both
cavities and qubits become entangled via Landau-Zener sweeps at different 
sweep velocities  $v_{A}$ and $v_{B}$, while the ancilla is in the
``off'' state, $g_{{\rm SW}} {=}\,0$. 
After the Landau-Zener sweeps, the state of the total system is
the product state
\begin{equation}
\label{rhoAB}
\rho_{\rm cav,i} (t)= \rho_{{\rm qA-cA}} \otimes \rho_{{\rm qB-cB}} ,
\end{equation}
where $\rho_{{\rm {q,A/B}-c {q,A/B}}}$ are of the form (\ref{rhoopen}).
At this point, there is no entanglement between any object of $A$ and
$B$. Once the state (\ref{rhoAB}) is achieved, the interaction is
``turned on'' towards a finite interaction strength $g_{{\rm
    SW}}\,{>}\,0$ at time $t\,{=}\,\tau_{\rm on}$. This may be
accomplished e.g.\ by changing adiabatically the operational point of
the ancilla, which modifies $\Delta$ \cite{Mariantoni2008a}. In the
following, correlations between the 
cavities $A$ and $B$ emerge by virtue of the interaction (\ref{Hint}).
At time $t\,{=}\,\tau_{\rm off}$, the interaction is turned off again. 
As far as we are only interested in the entanglement dynamics of the 
cavities, we may trace out the qubit and ancilla degrees of freedom in 
(\ref{rhoAB}). For $ \tau_\mathrm{off} -  \tau_\mathrm{on} = 2 \pi/
g_{{\rm SW}}$ the resulting density matrix $\rho_{{\rm cav,f}}$
takes the form  
\begin{eqnarray}
\label{rhocav1}
\rho_{\rm cav,f} = \left (
\begin{array}{cccc}
\overline \alpha & 0 & 0 & 0\\[.7ex]
0 & \frac{1}{2} (z_1+z_2) & \frac{-i}{2} (z_1-z_2)  & 0\\[.7ex]
0 & \frac{i}{2} (z_1-z_2) & \frac{1}{2} (z_1+z_2)  & 0\\[.7ex]
0 &  0 & 0 & \overline \beta \\
\end{array}
\right )
\end{eqnarray}
in the basis $\{  \ket{0_A 0_B}, \ket{0_A 1_B}, \ket{1_A 0_B}, \ket{1_A
  1_B}\}$. For simplicity, we have introduced the notations
\begin{eqnarray}\label{coeffs1}
\nonumber \overline \alpha & = & a_A a_B + a_A z_B + a_B z_A + z_A z_B \\
\nonumber \overline \beta  & = & d_A d_B \\
\nonumber z_1 & = & (z_A + a_A) \, d_B\\
z_2 & = & (z_B + a_B) \, d_A \, .
\end{eqnarray}
The coefficients  $a_{A/B}, d_{A/B}$ and $z_{A/B}$ are defined in 
Eq.~(\ref{coeffs}), where the indices $A$ and $B$ refer to the respective 
cavities. This reveals the dependence of the final state $\rho_{\rm
  cav,f}$ on the specific parameters of our particular entanglement
protocol. We point out that the most relevant parameters for the
preparation of  (\ref{rhocav1}) are the Landau-Zener sweep
velocities $v_A$ and $v_B$.  
The corresponding concurrence is given by
\begin{equation}\hspace{-2em}
C_\mathrm{cav,f}= 2 \,{\rm max} \left\{ 
0, \frac{1}{2} \left|z_1-z_2\right| - \sqrt{\overline \alpha \overline
  \beta} 
\right\}
\end{equation}
As a result, using the state preparation via Landau-Zener sweeps together with
the quantum switching protocol described above, we end up with both
cavities being in an entangled state that depends on the
particular parameters of the protocol. The individual entanglement
between the individual qubit-oscillator systems has thus been
\textit{transferred} via  cavity-cavity entanglement. 
%
%

\section{Finite-time disentanglement}
\label{suddendeath}

In the presence of a dissipative environment of each cavity,
the entanglement between the two cavities is naturally subject to
decay. Now the possibility of finite-time disentanglement in a quantum
optical setup due to spontaneous emission has been reported  in
Refs.~\cite{Yu2002a,Paz2008} for the special class of entangled two-cavity
states such as $\rho_{\rm cav,f}$; see Eq.~(\ref{rhocav1}). In this case
the concurrence $C(t)$  decays within a finite time, depending
on the initial state and thus on the particular state
preparation. This signifies a ``sudden death of (non-local)
entanglement'' whereas the decay characteristics of local decoherence 
is exponential. Finite-time disentanglement has also been
observed in a recent experiment \cite{Almeida2007}. 
\begin{figure}[tb]
  \centerline{\includegraphics[scale=.75]{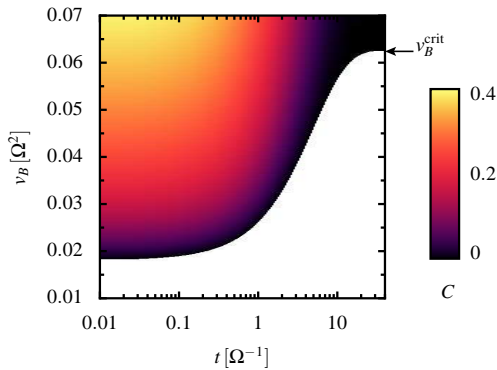}}
\caption{\label{fig:suddendeath}  (Color online) Time evolution of the
  concurrence   $C_\mathrm{cav} (t)$ starting with state 
  $\rho_\mathrm{cav,v}$. Here, $v_A=0.005\Omega^2, g_A=g_B=0.04\Omega$
  and   $T=0.001\Omega$. White color corresponds to separable
  states. Here, $v_B^\mathrm{crit}=0.0624 \Omega^2$.}
\end{figure}

In the following we study the disentanglement of the
state $\rho_\mathrm{cav,f}$. Again, as an important point, our
proposed circuit QED setup and protocol would allow for this to be
analyzed in the solid state realm, while Refs.~\cite{Yu2002a,Paz2008} refer
to an optical cavity QED setup.  

We start with the effective Hamiltonian  for the uncoupled cavities  
\begin{equation}
  \label{eq:Hcav-eff}
  \mathcal H _\mathrm{cav,f} ^\mathrm{eff} 
  = \hbar\Omega ( a_A^{\dagger} a_A ^{\vphantom{\dagger}} 
  + a_B^{\dagger} a_B ^{\vphantom{\dagger}} ) .
\end{equation}
Following section~\ref{entanglement-lzsweep}, we may restrict the
description of the cavities to the two lowest Fock states.
In order to investigate the time evolution of the entangled
two-cavity state $\rho_{\rm cav,f}$  after the ancilla interaction 
has been set to zero at $t=\tau_{\rm  off}$, we solve the quantum
master equation 
\begin{equation}
  \label{eq:QMEAB}
  \dot \rho =  -\frac{i}{\hbar}[ \mathcal H _\mathrm{cav,f}
  ^\mathrm{eff} ,\rho ] + \big(\mathcal L_A ( a_A
  ^{\vphantom{\dagger}} + a_A ^{\dagger}) + \mathcal L_B ( a_B
  ^{\vphantom{\dagger}} + a_B ^{\dagger}) \big) \,   \rho  
\end{equation}
where the Liouvillian $\mathcal L_{A/B}$ is defined as in
Eq.~(\ref{QME-cqed}). Hence we assume both entangled cavities to interact
individually with vacuum noise, i.e. to interact with individual
environments. A scenario with two two-level systems with zero
distance, i.e. coupled to the same environment, was discussed in
Ref.~\cite{Doll2006a}.

In Fig.~\ref{fig:suddendeath} we plot the concurrence $C_\mathrm{cav}
(t)$ for the decaying state $\rho_{\rm cav,f}$ as a function of time
and the sweep velocity $v_B$, while keeping $v_A$ constant. We find that
the concurrence drops to zero at a finite time $t \geq t_\mathrm{d}$
for a broad range of different values of $v_B$. This indicates that the
corresponding initial states become disentangled after a
finite time. However, this does not always hold:
For $v_B > v_B ^{\mathrm{crit}}$ we rather find exponential
concurrence decay, i.e. $t_\mathrm{d} = \infty$. Note that, in
contrast to these results, the qubit-cavity entanglement always decays  
exponentially as discussed in Section~\ref{dissipative}.

Our results comply perfectly with the disentangling
behavior discussed by Yu and Eberly \cite{Yu2002a}, which underlines
the generic nature of that scenario.


\section{Conclusions}

We have proposed a protocol for creating an entangled two-cavity state
within a circuit QED network architecture. Thereby,
we have considered a network of two superconducting qubits coupled
each to an electromagnetic quantum circuit. While there is no direct
interaction between the qubits, the cavities are connected through a
third ``ancilla'' qubit. Depending on its state and its detuning from
the cavities, this coupling may be switched on and off.
As first protocol step, one has to prepare two entangled
qubit-cavity states, which we have suggested to be done via
Landau-Zener sweeps. Note that there exist other possibilities for
this purpose, for example by Rabi oscillations. Second, the
interaction is ``switched on'' in order to entangle both cavities. At
the same time  we have considered the influence of  a separate
dissipative environment for each cavity.

A major goal here is the preparation of an entangled two-cavity state
such as (\ref{rhocav1}). In the quantum optics community this class of
states has been proposed to investigate finite-time
disentanglement due to spontaneous emission. Here we have shown that
this can be done as well in the solid state realm, depending on the
particulars of state preparation. The entries of the density matrix 
(\ref{rhocav1}) can be obtained by measuring cross-correlations
between the cavities. A corresponding circuit QED experiment has been  
proposed recently \cite{Mariantoni2006a}. 
Thus, we belief that our protocol will enable experimentalists to study
non-trivial dynamics of entanglement in quantum circuits.

\section*{Acknowledgements}

We thank Frank Deppe and Matteo Mariantoni for fruitful discussions.
This work has been supported by DFG through SFB 631 and by the German
Excellence Initiative via ``Nanosystems Initiative Munich (NIM)''.


\end{document}